\documentclass[reprint,superscriptaddress,
amsmath,amssymb,
aps,
pra, nobibnotes, nofootinbib
]{revtex4-2}

\usepackage{qip}
\usepackage{graphicx}
\usepackage{titling}
\usepackage[colorlinks=true,citecolor=red,urlcolor=blue,linkcolor=blue]{hyperref}
\usepackage{amsfonts}
\usepackage{amsmath,amssymb}
\usepackage{dsfont}
\usepackage{euscript}
\usepackage{float}
\usepackage{amsthm}
\usepackage{mathalfa}
\usepackage{bbm}
\usepackage{ragged2e}
\usepackage{color}
\usepackage{xcolor}
\usepackage{pgf}
\usepackage{color,soul}
\usepackage{tikz}
\usetikzlibrary{backgrounds}
\usepackage{float}
\usepackage{multirow}
\usepackage{array}
\usepackage{mathtools}
\usepackage[utf8]{inputenc}
\usepackage{lipsum}

\usepackage{array}
\usepackage{booktabs}
\usepackage{tabularx}
\usepackage{pifont}
\newcommand{\tr}{\mathrm{tr}}
\newcommand{\Tr}{\mathrm{Tr}}
\newcommand{\vb}[1]{\mathbf{#1}}

\newcommand{\sA}{\mathsf{A}}
\newcommand{\sB}{\mathsf{B}}
\newcommand{\sC}{\mathsf{C}}

\newcommand{\sE}{\mathsf{E}}

\newtheorem{df}{Definition}
\newtheorem{thm}{Theorem}

\theoremstyle{remark}

\newcounter{protcounter}
\newenvironment{protocol}[1]
    {\par \addvspace{\topsep}
        \noindent
        \tabularx{\linewidth}{@{} X @{}}
        \hline
        \refstepcounter{protcounter}
        \textbf{Protocol~\theprotcounter:} #1\\
        \hline}
    {\\
    \hline
    \endtabularx
    \par \addvspace{\topsep}}

\begin{document}
\title{Device-Independent Private Quantum Randomness Beacon}
\author{Ignatius William Primaatmaja}
\email{ign.william@gmail.com}
\affiliation{Squareroot8 Technologies Pte. Ltd., Singapore}
\author{Hong Jie Ng}
\affiliation{Squareroot8 Technologies Pte. Ltd., Singapore}
\author{Koon Tong Goh}
\affiliation{Squareroot8 Technologies Pte. Ltd., Singapore}

\begin{abstract}
    Device-independent quantum random number generation (DIQRNG) is the gold standard for generating truly random numbers, as it can produce certifiably random numbers from untrusted devices. However, the stringent device requirements of traditional DIQRNG protocols have limited their practical applications. Here, we introduce \textit{Device-Independent Private Quantum Randomness Beacon} (DIPQRB), a novel approach to generate random numbers from untrusted devices based on routed Bell tests. This method significantly relaxes the device requirements, enabling a more practical way of generating randomness from untrusted devices. By distributing the device requirements across a network of servers and clients, our proposal allows the server to operate high-performance devices while the clients can be equipped with more cost-effective devices. Moreover, the outputs of the client's device are also private, even against the server, which is essential in cryptographic applications. Therefore, DIPQRB provides a cost-effective method to generate secure and private random numbers from untrusted devices.
\end{abstract}
\maketitle

\section{Introduction}

The prospects of quantum advantage in secure communication~\cite{xu2020secure} and distributed computing~\cite{kimble2008quantum} have fueled the implementation of quantum networks~\cite{daiss2021quantum, pompili2021realization, chen_integrated_2021, van2022entangling, krutyanskiy2023entanglement, knaut2024entanglement} across the globe in recency. 
These advantages arise from the users' ability to detect adversaries and share quantum resources on the network, features that are typically elusive in classical networks. Until now, quantum networks have been used to demonstrate key distribution~\cite{chen_integrated_2021}, quantum-logic gate operations~\cite{daiss2021quantum, pompili2021realization}, and entanglement of qubits~\cite{van2022entangling, krutyanskiy2023entanglement, knaut2024entanglement}. We introduce a new use-case of quantum networks by showing that quantum networks can be used to generate secure and private random numbers in a cost-efficient manner.



Random numbers are a fundamental resource for modern computing and communication systems. They are vital for numerous essential applications that include cryptography, auditing, simulations, statistical analysis, and even blockchain~\cite{amer2025applications}. In particular, a Random Number Generator (RNG) is a critical component of a Hardware Root of Trust, which ensures the security of its cryptographic system's keys. As such, any open vulnerability of these RNGs may have dire consequences.

Owing to the probabilistic nature of quantum measurements, Quantum RNGs (QRNGs)~\cite{ma2016quantum}
produce random numbers that are provably secure. In contrast, outputs from RNGs that admit a classical description can be predicted with sufficient information and computation resources. 
Yet, there exist attacks in which malicious actors gain control over the QRNG output (for a concrete example, we refer the reader to Ref.~\cite{smith2021out}). 
They could do so by changing the quantum-state preparation and measurement processes, thereby invalidating the security proof of the QRNG.

By exploiting the phenomenon of Bell non-locality~\cite{bell1964, brunner2014bell}, one is able to guarantee the security of the output from a Device-Independent QRNG (DIQRNG)~\cite{colbeck2009quantum} without any assumption made on the inner-workings of the devices. Thus, DIQRNG is immune to such attacks by design. However, its implementations~\cite{pironio_random_2010, liu2018high, liu_device-independent_2021, shalm2021device} demand apparatus to meet highly stringent benchmarks, requiring DIQRNG users to possess expensive large form-factor devices, which renders DIQRNG impractical.

In this paper, we identify three desirable properties of QRNGs that are not simultaneously attainable with existing protocols. These desirable properties are (i) that the security of its output can be verified device-independently, (ii) that the QRNG users' devices can be small form-factor and cost-effective and (iii) that its output must not be correlated with registers possessed by the QRNG provider. The third requirement is especially important in cryptographic applications where the output random numbers are used as the users' secret keys. This property will ensure that nobody, not even the QRNG provider, will be able to guess the users' secret keys.

Existing commercial hardware QRNGs would fulfil (ii) and (iii), but not (i). DIQRNG implementations~\cite{pironio_random_2010, liu2018high, liu_device-independent_2021, shalm2021device} would fulfill (i) and (iii), but not (ii). The existing commercial QRNG on the cloud~\cite{huang2021quantum, kumar2022quantum} would fulfil (ii) but not (i) and (iii). In fact, Ref.~\cite{tamura2021quantum} reported that the random numbers generated by the authors on IBM 20Q Poughkeepsie did not pass the statistical test of NIST SP 800-22~\cite{SP800_22}, which demonstrates the danger of relying on device characterisation provided by an external vendor. The recent demonstration on certified quantum randomness based on quantum supremacy~\cite{liu2025certified} would fulfil (ii) and to some extent, (i), but not (iii). Interestingly, one may also deliver DIQRNG output via cloud but it would still not fulfil (iii). In order to fulfil all three properties, we introduce the \textit{Device-Independent Private Quantum Randomness Beacon} (DIPQRB).

For the conventional randomness beacon, the server broadcasts random numbers that are publicly available to its clients. In contrast, DIPQRB clients obtain private randomness from the quantum and classical information that the server broadcasts. On the other hand, DIPQRB enjoys the same economy-of-scale as its conventional counterpart by committing its resource-intensive processes in the server that serves multiple clients.

\begin{figure}
    \centering
    \includegraphics[width=\linewidth]{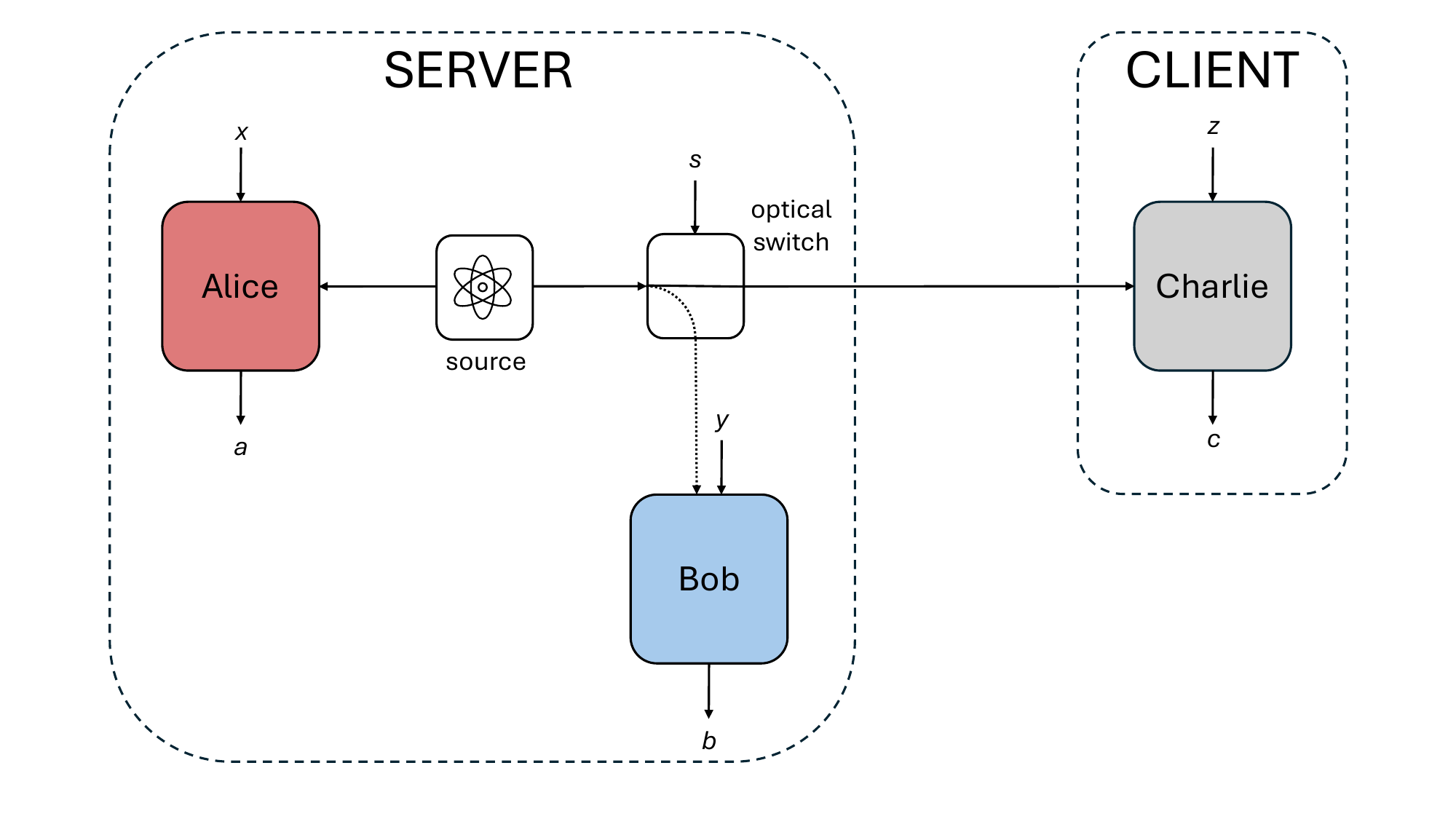}
    \caption{An illustration of a routed Bell test between three parties. The source emits a pair of entangled quantum states and sends one part to Alice. On the other side, an optical switch randomly routes the other part of the entangled state to either Bob or Charlie. In our proposal, Alice and Bob aims to establish a nonlocal correlation whereas Alice does not necessarily share a nonlocal correlation with Charlie, as long as they share the so-called long-range quantum correlation.}
    \label{fig: routed Bell test}
\end{figure}


The DIPQRB utilises the so-called routed Bell test configuration~\cite{lobo2024certifying} as illustrated in Fig.~\ref{fig: routed Bell test}. Unlike the typical Bell test, we have a router (e.g., an optical switch) that randomly distributes the entangled states between different parties that are involved in the routed Bell test. Furthermore, in the routed Bell test configuration, not all parties need to share non-local correlations to prove quantumness. This will reduce the requirements on the devices of some of the parties.

\begin{figure}
    \centering
    \includegraphics[width=\linewidth]{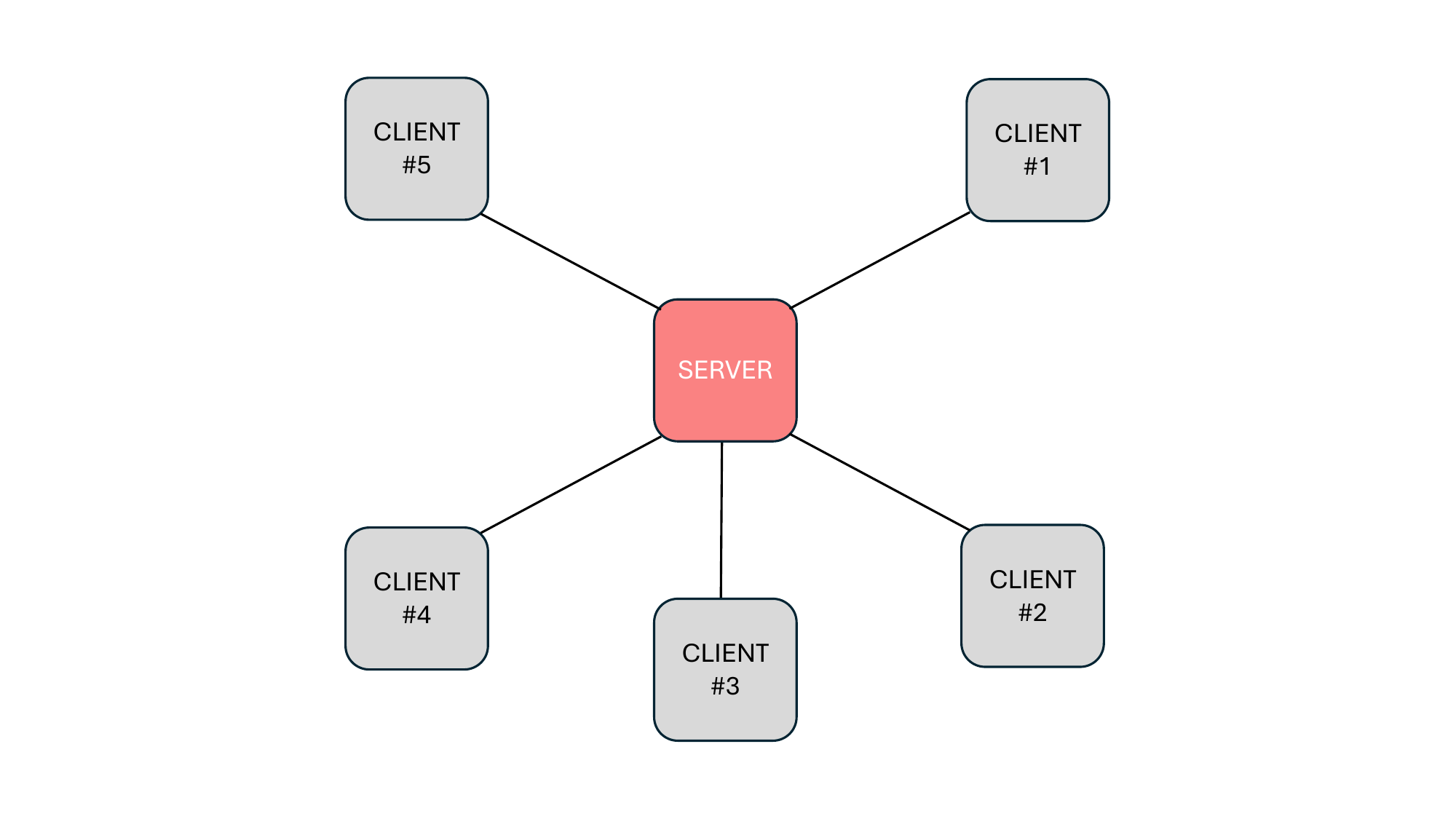}
    \caption{A network of a server and multiple clients. Only the server's devices are required to share a nonlocal correlation, which allows the clients to use detectors with lower quantum efficiencies. Since a single server can serve many clients, in a given network, the number of detectors with high quantum efficiencies will be much less than the ones with lower efficiencies.}
    \label{fig: network}
\end{figure}

We then envision a network comprising a server and multiple clients performing the routed Bell test (as shown in Fig.~\ref{fig: network}). In our proposal, the server, which contains multiple measurement devices, will perform a Bell test to produce a non-local correlation. For concreteness, we shall consider the case where there are two measurement devices inside the server, which we shall personify as Alice and Bob. The server will also randomly route one part of the entangled quantum state to a client, which will perform a random quantum measurement on it. We shall personify the client's measurement device as Charlie. After performing the routed Bell test for many rounds, the server will announce the inputs and outputs of Alice and Bob, as well as the setting of the router to the client. By comparing the server's announcement to Charlie's input and output data, the client can then quantify the randomness of Charlie's outputs.

Crucially, as the Bell test is only performed inside the server, the clients do not need to share non-local correlation with the server. Consequently, the clients do not need detectors that are extremely efficient. Indeed, it was recently showed it is possible to exhibit long-range quantum correlations in a routed Bell test even when the loss in the client's detector is arbitrarily large~\cite{sekatski2025certification}. Most of the cost is concentrated on the server's side, where a sophisticated setup is required to perform the Bell test. Furthermore, since one server can serve many clients, the cost of the setup required for the Bell test is shared among all the clients in the network. 

With sufficiently many clients, the average cost of implementing the network will be lowered due to economies-of-scale. Thus, our proposal will be suitable as a scheme for randomness-generation-as-a-service. Additionally, in contrast to randomness beacons, our scheme can generate \textit{private} randomness since the clients can generate random numbers that are uncorrelated to the server's data.

Our proposed scheme requires a quantum channel to connect the server to each client. This quantum channel can be realised using an optical fibre or a line-of-sight free space quantum link. Remarkably, the scenario that we consider here bears some resemblance to the scenario that is often encountered in quantum key distribution (QKD)~\cite{xu2020secure}, where an adversary can access the quantum channel to gain some side-information. In this case, it is imperative to analyse the security of our scheme against adversaries that may have quantum side-information. This is different from most of the existing QRNG schemes where the adversary is unable to access the quantum channel. In those cases, there is some flexibility to consider the nature of side-information that is accessible to the adversary.

\section{Protocol}
Our proposed protocol is as follows
\begin{protocol}{DI Private Quantum Randomness Beacon}
\textit{Protocol parameters}:
    \begin{itemize}
        \item $n$: number of rounds
        \item $\cX, \cY, \cZ$: the input alphabet of Alice, Bob, Charlie, respectively
        \item $\cA, \cB, \cC$: the output alphabet of Alice, Bob and Charlie, respectively
        \item $\varepsilon$: smoothing parameter
        \item $h$: the smooth min-entropy threshold
    \end{itemize}
\textit{Protocol}: 
\begin{enumerate}
    \item \textit{Routed Bell test}: For every round $i \in \{1,...,n\}$, repeat the following steps
    \begin{enumerate}
        \item \textit{Entanglement distribution}: A quantum source emits a pair of entangled quantum states. One part of the entangled state is sent to Alice, while the other part is sent to the optical switch.
        \item \textit{Routing}: The switch receives the input $S_i \in \{0,1\}.$ If $S_i = 0$, the switch directs the quantum state to Bob. If $S_i = 1$, then the switch directs the quantum state to Charlie.
        \item \textit{Measurement}: Alice receives the input $X_i \in \cX$, Bob receives the input $Y_i \in \cY$ and Charlie receives the input $Z_i \in \cZ$. Each party sets their measurement setting based on the input and then performs the quantum measurement. Alice records the output $A_i \in \cA$, Bob records the output $B_i \in \cB$ and Charlie records the output $C_i \in \cC$. If $S_i = 1$, Bob sets $B_i = \varnothing$ regardless of his original output.
        \item \textit{Server's announcement}: The server announces the data $P_i = (S_i, X_i, Y_i, A_i, B_i)$ to the client. If $S_i = 0$, Charlie sets $C_i = \varnothing$ regardless of his original output.
    \end{enumerate}
    \item \textit{Entropy estimation}: Based on the input-output data accumulated over $n$ rounds, the client estimates the conditional smooth min-entropy $H_{\min}^{\varepsilon}(\vb{C}|\vb{Z}, \vb{P}, E)$ and checks whether it is above a certain threshold $h$~\footnote{This is typically done by setting some range of accepted input-output frequency distributions, then checking if the observed frequency distribution lies within the range}. If yes, the client proceeds to the next step. Otherwise, the client will abort the protocol. Here, $E$ denotes the quantum side-information that any external party may have about the client's random string $\vb{C}$.
    \item \textit{Randomness extraction}: The client perform a randomness extraction protocol on the raw string $\vb{C}$ to obtain a uniformly random bit string.
\end{enumerate}
\end{protocol}
When both the locality and the detection loopholes are closed, the above protocol will be fully device-independent. In this case, the server's detectors need to have very high efficiencies and the source must be close to ideal. However, one can adopt a semi-device-independent (semi-DI) approach with some reasonable assumptions to reduce the requirements on the server's devices.

For example, note that the source, the optical switch, Alice, and Bob are all located inside the server. One can reasonably assume that the server is a secure location which makes it difficult for a potential adversary to physically access. Moreover, the server does not expect to receive any quantum signals from the client. Therefore, one can also use an optical power limiter~\cite{zhang2021securing} and an optical isolator to prevent attacks such as the detector blinding attack~\cite{lydersen2010hacking} from being launched via the quantum channel. One can then justifiably assume fair sampling, which means that we can safely discard the rounds where Alice or Bob do not produce a conclusive outcome.

Therefore, the same set of devices can perform both the fully device-independent QRNG protocol or the semi-DI protocol. This grants users the flexibility to operate the devices depending on their security needs. When the user is sceptical about the servers' devices, the fully device-independent mode of operation offers the highest level of security as all the devices can be fully untrusted -- as long as the server operates them honestly (e.g., it announces its inputs and outputs honestly during the protocol). This gold standard of security comes at the cost of lower randomness generation rate. On the other hand, if the user sufficiently trusts the server's devices, applying the fair sampling assumption on the server's measurement devices allows for higher randomness generation rate.

\section{Example}
As a simple example, we consider a particular protocol where each device has binary inputs, i.e., $\cX = \cY = \cZ = \{0,1\}$. In the semi-DI version of the protocol, we consider the case where each device has ternary outputs, i.e., $\cA = \cB = \cC = \{0,1,\varnothing\}$. The symbol $\varnothing$ corresponds to the case where the device does not produce a conclusive outcome. The fair sampling assumption imposes that in any given round, the probability of Alice (or Bob) obtaining the output $\varnothing$ is independent of their inputs
. We suppose that the client only generates random numbers in rounds where Alice produces a conclusive outcome (i.e., Alice does not produce the outcome $\varnothing$).

One can easily construct a fully device-independent version of the protocol by mapping the inconclusive outcome $\varnothing$ deterministically to the bit value `0'. Thus, in the fully device-independent version of the protocol, each party's output will also be binary, i.e., $\cA = \cB = \cC = \{0,1\}$. As the semi-DI version of the protocol offers a higher randomness generation rate without significantly reducing the level of security since the servers' devices are inside a secure location, we shall focus on this case. We can consider the case where Alice and Bob perform the Clauser-Horne-Shimony-Holt (CHSH) Bell test~\cite{CHSH} inside the server. In the CHSH Bell test, Alice and Bob estimate the so-called \textit{winning probability} $\omega$, defined as $
    \omega = \Pr[A_i \oplus B_i = X_i \cdot Y_i | S_i = 0, A_i \neq \varnothing, B_i \neq \varnothing]$,
where $\oplus$ here denotes addition modulo 2.

On the other hand, to test for long-range quantum correlations, Alice and Charlie will monitor the error rate $Q_0$ between Alice's and Charlie's first measurement as well as the error rate $Q_1$ between Alice's and Charlie's second measurement. The error rates are defined as
\begin{align*}
    Q_0 = \Pr[A_i \neq C_i| S_i = 1, X_i = Z_i = 0, A_i \neq \varnothing, C_i \neq \varnothing],\\
    Q_1 = \Pr[A_i \neq C_i| S_i = 1, X_i = Z_i = 1, A_i \neq \varnothing, C_i \neq \varnothing],
\end{align*}
The client also monitors the transmittivity $\tau_{z} = \Pr[C_i \neq \varnothing|S_i = 1, Z_i = z]$ for all $z$.

To illustrate the performance and robustness of the protocol, we consider an implementation with a spontaneous parametric down-conversion (SPDC) source and threshold single-photon detectors. We assume that Charlie's detector has an efficiency of $\eta$, and we assume that all the detectors have negligible dark count rates. We assume that the SPDC source is arranged such that the single-photon pair component is given by $
    \ket{\Psi_1} = \frac{1}{\sqrt{2}} \left(\ket{HH} + \ket{VV}\right)$,
i.e., the single-photon pair component gives a pair of photons that are maximally entangled in the polarisation degree of freedom. Here, $\ket{H}$ describes the horizontally polarised single-photon state, while $\ket{V}$ describes the vertically polarised single-photon state. We note that the SPDC source also has some vacuum and multi-photon components. For simplicity, we assume that the intensity of the source is set to be low enough that we can neglect the multiphoton components.

For the measurement, we assume that each measurement device consists of a pair of threshold single-photon detectors, a polarising beam-splitter and a polarisation modulator which decides the measurement basis. For our simulation, we set Alice's measurement basis to correspond to a polarisation rotation of $0^\circ$ for the setting $x = 0$ and $45^\circ$ for the setting $x = 1$. For Bob's measurement, the polarisation modulator will rotate the input polarisation by $22.5^\circ$ for $y = 0$ and $-22.5^\circ$ for $y=1$. Lastly, for the client's device, the polarisation modulator rotates the input polarisation by $0^\circ$ for $z = 0$ and by $45^\circ$ for $z = 1$, similar to Alice's device.

To simplify further, we consider the asymptotic case where the number of rounds are large, i.e., $n \rightarrow \infty$. In this case, we can take the limit $\varepsilon \rightarrow 0$. The randomness generation rate can be approximated by the single-round conditional von Neumann entropy $H(C_i|Z_i, P_i, E)$. In the practical case with finite number of rounds $n$, there will be a correction term that scales with $O(1/\sqrt{n})$~\cite{metger2024generalised}. This correction term vanishes in the asymptotic limit. More details are given in the Methods section.

\begin{figure}[h]
    \centering
    \includegraphics[width=\columnwidth]{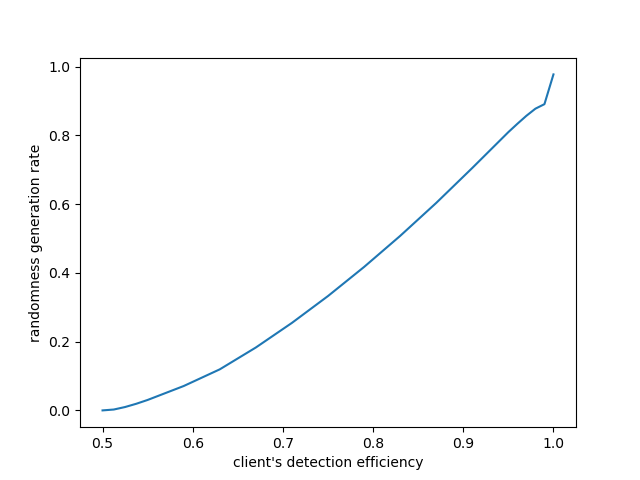}
    \caption{The plot of the asymptotic randomness generation rate per heralded event against the client's detection efficiency. Remarkably, the asymptotic randomness generation rate is positive as long as the client's detection efficiency is larger than 50\%.}
    \label{fig: rate}
\end{figure}

The randomness generation rate per heralded event (i.e., the events in which Alice does not produce the outcome $\varnothing$) is shown in Fig.~\ref{fig: rate}. We can see that the randomness generation rate is positive as long as the detection efficiency of the client's device is larger than 50\%. This is significantly lower than the detection efficiency required to perform the usual DIQRNG which is based solely on the CHSH game.

\section{Discussion and conclusion}
We note that the number of measurement bases on the client's side can be increased to improve the loss tolerance of the protocol~\cite{sekatski2025certification}. Remarkably, increasing the number of measurement bases of the client is relatively easy and the cost is relatively low, which further supports the claim that our solution is cost-effective. In particular, this means DIPQRB with more measurement bases for the clients can be implemented with single-photon detectors based on avalanche photodiodes instead of superconducting nanowire single-photon detectors that are significantly more expensive and need to operate at cryogenic temperatures. 

In conclusion, DIPQRB is a promising near-term application of quantum networks. More concretely, it enables the generation of certifiable private randomness in a cost-effective way from untrusted devices. With the ongoing efforts to build quantum networks around the world, we foresee that DIPQRB will play a vital role in applications where secure and private randomness are indispensable. Furthermore, since the same quantum devices and infrastructure that are used in DIPQRB can also be used for other applications, such as QKD, this will further justify the investment in building these quantum networks.

\section{Methods}
We now present the method for entropy estimation for the protocol presented in the main text. As we mentioned in the footnote, in practice, the entropy estimation is typically done by checking if the observed input-output frequency distribution is within some accepted range. We denote the range of accepted distributions as $\cS_{\Omega}$. In this section, we want to show that if the observed input-output frequency distribution $\vec{q}$ is within the accepted range, i.e., $\vec{q} \in \cS_{\Omega}$, then the conditional smooth min-entropy is higher than some threshold, except with a small probability. In this work, we use the conditional smooth min-entropy as a measure of randomness since it quantifies the amount of private random bits that the client can extract from the protocol.

\subsection{Assumptions}
Before we elaborate on the entropy estimation method further, we shall list our assumptions:
\begin{enumerate}
    \item Quantum theory is correct.
    \item The inputs for the server's and the client's devices are chosen from trusted random sources that are independent from the adversary and from each other.
    \item The server and the client follow the protocol honestly. We say that the server is \textit{honest-but-curious}.
    \item The server and the client's devices do not communicate with one another, unless required by the protocol.
    \item The client's device does not leak any information to the adversary nor the server.
    \item (For semi-device-independent version of the protocol) Whether or not the server's devices (Alice and Bob) register a click in a given round, is independent of the inputs to the devices.
\end{enumerate}

The first assumption is necessary since our analysis is based on quantum theory, thus the security of the protocol relies heavily on the correctness of quantum theory. Next, we use the second assumption to greatly simplify the analysis. In recent years, there are new techniques to analyse the security of quantum cryptographic protocols with non-ideal sources of randomness, hence there is a possibility that this assumption can be relaxed. The third assumption is necessary since the client requires the inputs and outputs to the server's devices to estimate the entropy produced in the protocol. That being said, if the server's data is leaked after the protocol execution, we claim that the client's random string will still remain private, thus the server only needs to behave honestly during the protocol execution. The fourth assumption is a standard assumption in routed Bell tests~\cite{lobo2024certifying}, which form the basis of our protocol. Thus, our protocol naturally inherits this assumption. The fifth assumption is a standard assumption in cryptography. Finally, for the semi-device-independent version of the protocol, we would like to discard rounds where Alice does not register a click, and similarly we would like to discard rounds where the switch directs the photon to Bob and he does not register a click. This is only permissible if whether or not the device registers a click is independent of the input given to that device. We note that this assumption is not necessary in the fully-device-independent framework.

\subsection{Generalised entropy accumulation theorem}
To estimate the entropy that the protocol generates, we shall use the generalised entropy accumulation theorem (GEAT) with testing~\cite{metger2024generalised}. In a nutshell, GEAT is a powerful tool to estimate the entropy produced by a sequence of processes, which obey certain conditions. It can elegantly account for non-independent-and-identical attacks as well as memory effects in the quantum devices.  It has been used to analyse a wide-range of quantum cryptographic protocols, especially their finite-size effects. However, the study of the finite-size effects of our protocol is beyond the scope of the current work, and we mainly want to show (using GEAT) that our protocol remains secure even if the adversary launches a sophisticated attack that might depend on the data that she obtains in the preceding rounds, as well as when the devices' behaviours depend on the previous inputs and outputs. To that end, we want to show that our protocol satisfies the conditions required to apply the GEAT and that the limit of very large number of rounds, the randomness generation rate is lower bounded by the rate that we claim in the main text. We shall leave the finite-key analysis of our protocol for future work.

To apply the GEAT, we need to introduce a few notions
\begin{df}[GEAT channels]
    Let $\{\cM_i\}_{i=1}^n$ be a collection of completely positive, trace preserving (CPTP) maps with $\cM_i: R_{i-1}E_{i-1} \rightarrow C_i D_i R_i E_i$, where the register $D_i$ is classical. We say that $\{\cM_i\}_{i=1}^n$ are GEAT channels if
    
    \begin{enumerate}
        \item For all $i \in [n]$, the channel $\cM_i$ has the following non-signalling property: there exists a CPTP map $\cN_i: E_{i-1} \rightarrow E_i$ such that
        \begin{equation}
            \Tr_{R_{i} C_{i} D_i} \circ \cM_i = \cN_i \circ  \Tr_{R_{i-1}}
        \end{equation}
        \item Let $\cM'_i = \Tr_{D_i} \circ \cM_i$. There exists a channel $\cT: C^n E_n \rightarrow C^n E_n D^n$ of the following form
        \begin{multline}
            \cT[\omega_{C^n E_n}] = \sum_{p, q} \left(\Pi_{C^n}^{(p)} \otimes \Pi_{E_n}^{(q)}\right) \omega_{C^n E_n} \left(\Pi_{C^n}^{(p)} \otimes \Pi_{E_n}^{(q)}\right) \\
            \otimes \ketbra{F(p,q)}{F(p,q)}_{D^n},
        \end{multline}
        where $\{\Pi_{C^n}^{(p)}\}_{p}$ and $\{\Pi_{E_n}^{(q)}\}_{q}$ are mutually orthogonal projectors on $C^n$ and $E_n$, respectively, and $F$ is a deterministic function.
    \end{enumerate}
\end{df}

The GEAT channels describe the actions of the server, the client and the adversary in each round. After the $i$-th round, the registers $E_i$ and $R_i$ describe the side-information available to the adversary and the internal memory of the client's device, respectively. On the other hand, the register $C_i$ stores the client's measurement outcome in the $i$-th round. The classical register $D_i$ is used by the client to ``test'' the protocol. The non-signalling property (the first condition) can be interpreted as the client's device not leaking any secret information to the adversary -- an important property to prevent a trivial attack such as the client's device leaking the measurement outcomes covertly to the adversary.

In this work, we consider the so-called \textit{infrequent sampling channels}. These are GEAT channels that have the following additional structures. Consider a virtual protocol where the client randomly assigns each round $i$ to either test round or generation round with probability $\gamma$ and $1-\gamma$, respectively. In the generation round, the register $D_i$ is always mapped to $D_i = \perp$. In the test round, it will take a value from the alphabet $\cD \setminus \{\perp\}$, depending on the inputs and outputs of the server's and client's devices. In this virtual protocol, we shall consider the case where the client announces $D_i$ in each round -- thus, reducing the conditional entropy. Clearly, the actual protocol where the client does not reveal $D_i$ will produce higher conditional entropy. Therefore, estimating the entropy produced in the virtual protocol provides a conservative estimate of the entropy produced in the actual protocol.

The use of the virtual protocol is for the sake of simplicity of the entropy estimation. More concretely, we construct the virtual protocol so that we can directly use the GEAT with testing presented in Ref.~\cite{metger2024generalised}. We note that it is possible to avoid this virtual protocol construction by using private test registers (such as the one presented in Ref.~\cite{arqand2024generalized}), which will lead to higher certified conditional entropy. This is not our immediate concern since the finite-size effects of the protocol is not within the scope of this Letter, and hence we can set the probability  $\gamma$ of choosing a test round to be arbitrarily small.

Now, let us construct the GEAT channels for the protocol. For each $i$, each channel $\cM_i$ is constructed as follows:
\begin{enumerate}
    \item The adversary prepares the quantum state using the channel  $\cM^{\mathrm{prep}}_i: E_{i-1} \rightarrow Q_{A_i} E'_i$ based on her side information in the previous round. Then, the adversary sends the quantum register $Q_{A_i}$ to Alice and the quantum register $E'_i$ to the switch.
    \item The channel $\cM^{\mathrm{switch}}_i : E_i' \rightarrow Q_{B_i} Q_{C_i} E''_i$ randomly generates the input $S_i \in \{0,1\}$ to the switch, then depending on $S_i$ and $E'_i$, the switch prepares the quantum registers $Q_{B_i}, Q_{C_i},$ and $E''_i$. The quantum register $Q_{B_i}$ is sent to Bob, the quantum register $Q_{C_i}$ is sent to Charlie and the quantum register $E''_i$ is kept by the adversary.

    \item The channel $\cM^{\mathrm{meas}}_i: Q_{A_i} Q_{B_i} Q_{C_i} R_{i-1} S_i \rightarrow X_i A_i Q'_{A_i} Y_i B_i Q'_{B_i} Z_i C_i R_i S_i$ generates the input $X_i$ for Alice, then measures the register $Q_{A_i}$ to obtain the outcome $A_i$ and the post-measured quantum register $Q'_{A_i}$. Similarly, the channel generates the input $Y_i$ for Bob, then measures the register $Q_{B_i}$ to obtain the outcome $B_i$ and the post-measured quantum register $Q'_{B_i}$. The channel also generates the input $Z_i$ for Charlie, then measures the register $Q_{C_i}$ to obtain the outcome $C_i$ and the post-measured quantum register $R_i$. Lastly, if $S_i = 0$, Charlie sets $C_i = \varnothing$. Otherwise, if $S_i = 1$, Bob sets $B_i = \varnothing$.

    \item The channel $\cM^{\mathrm{test}}_i: S_i A_{i} B_i C_{i} X_i Y_i Z_i \rightarrow S_i T_i A_{i} B_i C_{i} X_i Y_i Z_i D_i$ generates the classical register $T_i \in \{0,1\}$ based on $S_i$\footnote{The probability of generating $T_i = 1$, conditioned on $S_i = 0$ may be different from respective probability conditioned on $S_i = 1$.}. The register $T_i$ denotes whether the $i$-th round is used as a test round or a generation round. Then, the channel generates $D_i$ based on $T_i$ and $S_i, A_i, B_i, C_i, X_i, Y_i, Z_i$.
    If $T_i = 0$, then set $D_i = \perp$. Otherwise, set $D_i = F(A_i, B_i, C_i, X_i, Y_i, Z_i, S_i)$ for some deterministic function $F$. 

    \item The channel $\cM_i^{\mathrm{collect}} : X_i Y_i Z_i S_i T_i A_i B_i Q'_{A_i} Q'_{B_i} E''_i \rightarrow E_i$ collects the registers $X_i, Y_i, Z_i, S_i, T_i, A_i, B_i,$ $Q'_{A_i}, Q'_{B_i}, E''_i$ and gives them to the adversary, forming the quantum register $E_i$.
\end{enumerate}
Then, we can construct the GEAT channel $\cM_i = \cM_i^{\mathrm{collect}} \circ \cM_i^{\mathrm{test}} \circ \cM_i^{\mathrm{meas}} \circ \cM_i^\mathrm{switch} \circ \cM_i^{\mathrm{prep}}$. Clearly, this resembles the interactions in the actual protocol.

Now, a few remarks are in order. First, the memory register $R_i$ is only associated with the client's device, Charlie. The server's devices' memories are collected by the adversary at the end of each round, via the channel $\cM_i^{\mathrm{collect}}$. This is to reflect the fact that we want the protocol to be secure even if an adversary has access to the server's devices. Furthermore, this also helps to satisfy the non-signalling property as the adversary wouldn't be able to update her side information about $A_i$ and $B_i$ without having access to the memories of Alice and Bob (which are stored in the quantum registers $Q'_{A_{i-1}}$ and $Q'_{B_{i-1}}$ and are also collected by $\cM_{i-1}^{\mathrm{collect}}$). Next, the construction of the GEAT channel also sheds some light on the level of trust that we need from the server. The GEAT channel construction requires the following conditions: (1) the preparation of the quantum register $Q_{A_i}$ must be independent of the input of the switch $S_i$, (2) the preparation of the quantum register $Q_{B_i}$ is independent of $X_i$ and $Z_i$, and similarly, the preparation of the quantum registers $Q_{C_i}$ is independent of $X_i$ and $Y_i$, and lastly, (3) the computation of $D_i$ has to be done honestly -- in practice, this requires the server to send the registers $A_i B_i X_i Y_i S_i$ honestly. The first two conditions are the standard non-signalling condition in routed Bell tests~\cite{lobo2024certifying}. Therefore, we require the server to behave honestly to respect the routed Bell test non-signalling condition as well as to allow the client to estimate the input-output distribution based on the actual inputs and outputs. Apart from these three conditions, the server's devices (and the client's device) may deviate from their expected behaviour.

Having introduced the GEAT channels and showed how the interactions in the protocol can be described as GEAT channels, we now introduce the notion of min-tradeoff functions
\begin{df}[Min-tradeoff functions]
    Let $\cD$ be an alphabet and let $\mathbb{P}_{\cD}$ be the set of probability distributions over $\cD$. The affine function $f: \mathbb{P}_{\cD} \rightarrow \mathbb{R}$ is called a min-tradeoff function for $\{\cM_i\}_{i=1}^n$ if
    \begin{equation}
        f(\vec{q}) \leq \min_{\nu \in \Sigma_i(\vec{q})} H(C_i|E_i, \tilde{E}_{i-1})_\nu,
    \end{equation}
    where the set $\Sigma_i(\vec{q})$ is defined as
    \begin{multline}
        \Sigma_i(\vec{q}) = \Bigg\{ \nu_{C_i R_i D_i E_i \tilde{E}_{i-1} } = (\cM_i \otimes \id_{\tilde{E}_{i-1}} ) [\sigma_{R_{i-1} E_{i-1} \tilde{E}_{i-1}}]: \\
        \sigma \in \mathrm{D}(R_{i-1}E_{i-1} \tilde{E}_{i-1}), \, \nu_{D_i} = \vec{q} \Bigg\}, 
    \end{multline}
    where $\mathrm{D}(R_{i-1} E_{i-1} \tilde{E}_{i-1})$ denotes the set of normalised quantum states over the registers $R_{i-1} E_{i-1} \tilde{E}_{i-1}$ and the register $\tilde{E}_{i-1}$ is isomorphic to the registers $R_{i-1} E_{i-1}$.
\end{df}

Informally, the min-tradeoff function (evaluated on the distribution $\vec{q}$) gives a lower bound on the single-round conditional von Neumann entropy evaluated on quantum states whose system $D_i$ is distributed according to $\vec{q}$. The technique to construct the min-tradeoff function has been given in Refs.~\cite{dupuis2019entropy, metger2024generalised}.

Now, let us recall the GEAT with testing derived in Ref.~\cite{metger2024generalised}
\begin{thm}[\textbf{Corollary 4.6} of Ref.~\cite{metger2024generalised}]
    Let $\Omega \subseteq \cD^n$ be an event and let $\cS_{\Omega}$ be the set of frequency distributions over $\cD$ that are compatible with the event $\Omega$. Furthermore, let $\rho_{C^n D^n E_n|\Omega}$ be the output of the GEAT channels conditioned on this event after tracing out $R_n$. Let $\varepsilon \in (0,1)$. Then,
    \begin{equation}
        H_{\min}^{\varepsilon}(C^n|D_n, E_n)_{\rho_{|\Omega}} \geq n h_{*} - O(\sqrt{n}),
    \end{equation}
    where $h_{*} = \min_{\vec{q} \in \cS_{\Omega}} f(\vec{q})$.
\end{thm}

Informally, GEAT states that up to the first order, the conditional smooth min-entropy conditioned on the event $\Omega$ is proportional to the number of rounds in the protocol, with the constant of proportionality $h_{*}$ given by the min-tradeoff function, evaluated on the worst case distribution that is compatible with the event $\Omega$. To apply this to our protocol, this event $\Omega$ is the event in which the protocol is not aborted, hence GEAT gives a lower bound on the conditional smooth min-entropy produced by the protocol.

\subsection{Single-round analysis}
The GEAT essentially reduces the analysis of the conditional smooth min-entropy for $n$ rounds of the protocol to the construction of a min-tradeoff function, which is related to the single-round conditional von Neumann entropy. To obtain the lower bound on the single-round conditional von Neumann entropy, we lower bound it using the conditional min-entropy, which is related to the guessing probability
\begin{align}
    &H(C_i|A_i, B_i, X_i, Y_i, S_i, E) \\&\geq \Pr[S_i = 1, Z_i \neq X_i] H(C_i|S_i = 1, Z_i \neq X_i, A_i E) \nonumber\\
    &\geq  \Pr[S_i = 1, Z_i \neq X_i] H_{\min}(C_i|S_i = 1, Z_i \neq X_i, A_i E)\\
    & = -\Pr[S_i = 1, Z_i \neq X_i] \log_2 \left(P_g(C_i|S_i = 1, Z_i \neq X_i, A_i E)\right)
\end{align}
The first inequality is due to us discarding the non-negative terms corresponding to $S_i = 0$ or $Z_i = X_i$. For $S_i = 1$ terms, the conditioning on $B_i$ is trivial since $B_i = \varnothing$. Similarly, the conditioning on $Y_i$ can be removed due to the non-signalling condition. The second inequality is due to the conditional min-entropy being a lower bound on the conditional von Neumann entropy and the last equality is due to the definition of the conditional min-entropy with $P_g$ denoting the guessing probability.

The guessing probability can then be bounded using the Navascu\'es-Pironio-Ac\'in (NPA) hierarchy~\cite{npa}. We shall solve the following optimisation problem
\begin{widetext}
\begin{equation}
\begin{split}
    P_g(C_i|Z_i \neq X_i, S_i = 1, A_i E)
    = \max &\sum_{x, z} \Pr[X_i = x, Z_i = z|S_i = 1] \sum_{a,c} \bra{\psi} \sA_{a|x} \sC_{c|z} \sE_{c|axz} \ket{\psi}\\
    \mathrm{s.t.} \, & \bra{\psi}\sA_{a|x} \sB_{b|y}\ket{\psi} = p(ab|xy, s=0), \qquad \forall a,b,x,y\\
    & \bra{\psi}\sA_{a|x} \sC_{c|z}\ket{\psi} = p(ac|xz,s=1), \qquad \forall a,c,x,z\\
    & [\sA_{a|x}, \sB_{b|y}] = 0, \qquad \forall a,b,x,y\\
    & [\sA_{a|x}, \sC_{c|z}] = 0, \qquad \forall a,c,x,z\\
    & [\sA_{a|x}, \sE_{e|a'x'z'}] = 0, \qquad \forall a,a',e,x,x',z'\\
    & [\sC_{c|z}, \sE_{e|a'x'z'}] = 0, \qquad \forall c,a',e,z,x',z',\\
    & \sA_{a|x} \sA_{a'|x} = \delta_{aa'} \sA_{a|x}, \qquad \forall x\\
    & \sB_{b|y} \sB_{b'|y} = \delta_{bb'} \sB_{b|y}, \qquad \forall y\\
    & \sC_{c|z} \sC_{c'|z} = \delta_{cc'} \sC_{c|z}, \qquad \forall z\\
    & \sE_{e|axz} \sE_{e'|axz} = \delta_{ee'} \sE_{e|axz}, \qquad \forall a,x,z
\end{split}
\end{equation}
\end{widetext}
Here, $\{\sA_{a|x}\}_{a,x}$, $\{\sB_{b|y}\}_{b,y}$, $\{\sC_{c|z}\}_{c,z}$, and $\{\sE_{e|axz}\}_{e,a,x,z}$ corresponds to the projectors for Alice, Bob, Charlie, and the adversary, respectively. We note that Bob's projectors may not commute with those of Charlie and the adversary since the behaviour of the switch is not trusted.

We note that it is also straightforward to modify the optimisation problems for protocols that monitor linear combinations of $\{p(ab|xy,s=0)\}_{abxy}$ and linear combinations of $\{p(ac|xz,s=1)\}_{acxz}$. From the guessing probability, we can obtain an affine min-tradeoff function by simply linearising the logarithmic function by taking a tangent at any given point in the interval $[0,1]$. We also note that it is possible to tighten our bound by using the technique presented in Refs.~\cite{tan2024entropy,deloison2025device} which is based on the Brown-Fawzi-Fawzi bound~\cite{brown2024device}, however, this will require us to solve a bigger SDP.
\bibliography{references}
\end{document}